\documentclass[12pt]{article}
\usepackage{subfigure}
\usepackage{epsfig}
\newcommand{\etal}{{\em et\/al\/.}\hspace{0.5em}}
\newcommand{\ie}{{\em i\/.e\/.} \hspace{0.5em}}

\newcommand{\eg}{{\em e\/.g\/.}\hspace{0.5em}}
\begin{document}
\title{Particle Candidates of Ultrahigh Energy Cosmic Rays\thanks{Review talk given at the 
Europhysics conference, HEP2001,
Budapest, Hungary.}}
\author{G.~Domokos and S.~Kovesi-Domokos\\
Department of Physics and Astronomy\\
The Johns Hopkins University\\
Baltimore, MD 21218\thanks{e-mail: skd@jhu.edu}}
\date{October  2001}
\maketitle
\abstract{We discuss candidates for trans-GZK cosmic rays observed in a variety of detectors.
Three types of primaries are represented among the abstracts submitted to this meeting:
neutrinos causing a Z-burst, protons arising from the decay of ultra-heavy metastable
particles and neutrinos within the framework of low scale string-like models of unification.
We attempt to evaluate the relative merits of these schemes. No definite conclusion can be 
reached at this time. However, we point out that some schemes are more credible/predictive
than others. Data to be gathered by  the Pierre Auger observatories as well as orbiting
detectors (OWL, Airwatch\ldots) should be able to decide between the various schemes.}
\newpage
\section{Introduction}
As soon as the cosmic microwave backround radiation (CMBR) was discovered, Greisen and
Zatsepin and Kuzmin independently pointed out that there should exist a cutoff in the 
primary cosmic ray spectrum~\cite{GZK}. The physical basis of that cutoff is
 very well established. In fact, the physical process responsible for it was originally
discovered by Enrico~Fermi and his group in the fifties and has been extensively studied 
ever since.
It is generally believed that the primary cosmic radiation at energies above, say,
$10^{17.5}$eV or so, consists mostly of protons. The composition at these energies 
is discussed by Stecker in his excellent Erice lectures~\cite{stecker}.  
Protons interact with the
CMBR, produce electron pairs,  photoproduce pions 
and lose about 30\% of their energy per interaction. (The production rate of more than 
one pion is negligibly small.)
This means that, unless protons
are produced in our ``cosmic backyard'', they cannot have energies exceeding
some $10^{19}$eV or so  (the so-called {\em GZK cutoff}\/). Else,
if produced far away with energies sufficiently high to allow them
to arrive here above the GZK cutoff, their interaction with the CMBR would ultimately produce
an unacceptably high  photon background at lower energies, see  Stecker,~{\em loc. cit.}
for a review.

At the time of this writing some $20+$ events have been observed of energies
exceeding the GZK cutoff and they have generated a flurry of theoretical papers
exceeding the number of events observed by at least a factor of two or more. (The
highest energy published event is still the one  reported by Fly's Eye, 
see~\cite{birdetal}.\footnote{Since the time this talk was given, another event of comparable or
somewhat higher energy was reported. See M.~Teshima, talk presented at the
{\em Summer study on the future of Particle Physics, Snowmass, CO 2001}}) 
 Clearly, it is unsatisfactory to have a small number of
events observed in the trans-GZK category. However, with the Pierre~Auger observatories,
OWL and Airwatch becoming operational within the near future, this situation will change
drastically.
\vspace{2truemm}
{\bf Should we get excited -- if so, why?}
\vspace{2truemm}

There is a variety of attitudes adopted in connection with the reported trans-GZK
events. Here is a sample based on our informal evaluation of a variety of papers and 
conversations with some colleagues.
\begin{itemize}
\item {\em Denial.} OK, some trans-GZK events have been reported. However, it is known that
the primary cosmic ray (CR) spectrum is rapidly falling at energies larger than, say,
$10^{17}$eV. In such a situation  it is easy to overestimate the
primary energy. Maybe, the problem will just go away with more events observed.
\item {\em Our backyard.} The GZK cutoff can be overcome if there is
a sufficient number of local sources of UHECR. Blanton \etal \cite{blanton}
made the reasonable  assumption that the distribution of sources follows
the distribution of galaxies. They find a local overdensity of $\approx 2$
and they conclude that for soft injection spectra this overdensity is
insufficient to explain the observed flux around $10^{20}$eV.

Despite the original analysis  of 
Elbert~and~Sommers~\cite{elbertsommers}, there may be {\em individual} sources of 
trans-GZK CR in our
galactic neighborhood. In particular, CEN~A has been conjectured as a possible source
of such trans-GZK protons~\cite{farrarpiran}. With a distance of about 3.5Mpc, CEN~A
``just barely makes it'' as a potential source. In contrast to older calculations, 
the thorough analysis of Stanev {\em et al}~\cite{stanevetal} results in a 
much smaller radius of the GZK sphere (about 14~Mpc or so) rather than 
about 50~Mpc as  believed 
previously\footnote{The GZK sphere is a somewhat fuzzy notion: roughly speaking,
it is the radius of the sphere within which a source has to lie in order to
provide us  with protons of $10^{20}$eV of energy. The work of Stanev {\em et al.}
contains precise definitions. The difference between this work and previous ones 
lies in a careful accounting of intergalactic magnetic fields, which either were not treated 
or only in some sketchy way. A  notable exception is 
ref.~\cite{achterberg}.}.
 
\vspace{1mm}
{\footnotesize  In order to illustrate  what is involved 
in ref.~ \cite{stanevetal},
consider the following simple model. Roughly speaking, a proton executes 
a random walk in the chaotic magnetic fields. Let $L$ denote the total path 
length necessary for the proton to lose a substantial fraction of its energy,
\ie the GZK length. Since the particle is ultrarelativistic, $t=L$ to a
high degree of accuracy. The rms  distance the proton travels in time $t$
is therefore given by
\[ \sqrt{ \langle {\bf x}^{2} \rangle } \approx \sqrt{\lambda t} \approx \sqrt{\lambda L}, \]
where $\lambda$ is the mfp of magnetic scattering; it is of the order
of a few Mpc. Assuming,  $\lambda \approx 5$Mpc,
and $L=50$Mpc, one 
gets $\sqrt{\langle {\bf x}^{2}\rangle} \approx 16 \mbox{\rm Mpc}$. 
Considering the crudeness of all the approximations made here, this is
in reasonable agreement with the result of ref.~\cite{stanevetal}.}

\vspace{1mm}
At present, it is not clear whether CEN~A can account for the
observed trans-GZK events. Like previous similar proposals, the CEN~A hypothesis 
has been criticized, see \eg ref.~\cite{isola}. In a sense, the controversy
revolves around the question whether there are sufficiently strong random 
intergalactic magnetic fields to  wash out  the directionality of the
UHECR arriving from a single source.
 An interesting test of the CEN~A hypothesis was proposed by 
Anchordoqui {\em et al.}, ref.~\cite{anchordoqui}.  When the southern Auger Observatory  
will be operational, one should look for neutrons emitted by CEN~A. At 
such energies, neutrons appear as stable particles. Being neutral, they are less affected by
the magnetic fields on the way here than protons are. 

Should the CEN~A hypothesis (or some other similar one, \eg M87 as a source) prove
viable, the problem of trans-GZK cosmic rays will be resolved without invoking 
new particle physics or substantially new astrophysics.
\item {\em New physics and/or astrophysics\/.} All works submitted to this meeting
dealing with the problem of trans-GZK cosmic rays belong to this category. 
They represent approaches which, so far, have not been definitely eliminated
either by accelerator based experiments ({\em e.\/g.\/} lower limits on the masses of
superpartners) or by other physical or astrophysical arguments.
\end{itemize}
If the first two items mentioned above get somehow eliminated, then our answer is
that {\em yes, we should get excited: the trans-GZK CR are messengers of some 
new particle physics or new astrophysics\/.}
In the following sections, we discuss these works.
\section{Z-bursts and Neutrino Clouds.}
The original idea of Z-bursts was conceived by Fargion \etal and by 
 Weiler,~\cite{weiler82}. The basic idea is
 beautifully simple  and it needs no particle physics beyond the well established
Standard Model.

The authors argue  that neutrinos can penetrate the CMBR, because they have  a very small 
magnetic moment and in  scattering on a CMBR photon, the CMS energy, $\sqrt{s}$ is about
100MeV or so. There may be a halo of neutrinos around our galaxy. If  neutrinos are
massive, of mass $m_{\nu}$, an interaction between a neutrino coming from afar and a halo (anti)
neutrino can excite the $Z$ resonance. The neutrino energy needed for that is
about $E= M_{Z}^{2}/2 m_{\nu}$.
The, $Z$, in turn, decays into nucleons, roughly 35\%
of the time and {\em voil\'{a}}, we have the trans-GZK cosmic rays. 
 Strictly speaking, the neutrinos in the halo need not be massive. For massless neutrinos, 
just replace  $m_{\nu}$ by $\approx kT$, with $T=1.9 ^{{\rm o}}K$. 
The trouble is
that one needs inconceivably high neutrino energies if the neutrino mass is
replaced by the thermal energy of a massless relic neutrino.

Waxman pointed out that the total energy carried by the high energy neutrino flux
was dangerously close to the total luminosity of the Universe~\cite{wax98}. Clustering 
or a high lepton asymmetry~\cite{gelminikusenko} is needed in order to make 
the Z-burst  scenario 
work. Unfortunately, a large lepton asymmetry leaves its imprint on the CMBR and the latest
observations appear to be incompatible with the asymmetry postulated by Gelmini and 
Kusenko~\cite{riotto}.

Now McKellar \etal~\cite{mckellar} in a paper submitted to this meeting  argue that 
clustering in the neutrino halo can be accomplished if one revives the idea of neutrino 
clouds~\cite{stephenson}. The model is based on  the assumption of  a light scalar field
weakly coupled to neutrinos. By performing  a self consistent field calculation, the authors
argue that the minimum of field energy is obtained if the neutrino beckground is not
uniform, but clustered. 
 This clustering phenomenon is somewhat similar to what is found in a realistic ferromagnet:
the minimum of the free energy is obtained by forming domains instead of a completely
homogeneously oriented spin array. This scenario has a chance to rescue the Z-burst model.

The authors consider a broad range of densities and cluster radii, within bounds imposed by
the amount of dark matter in the solar system and upper bounds on the incident neutrino flux.

There are several questions to be answered, however,  in this connection.
\begin{itemize}
\item Are there any other consequences? (For instance, imprint on the CMBR, cosmology, 
{\em etc\/.}) What is the cosmological density of the light scalar? Does it have an
effect on nucleosynthesis?
\item Several theoretical questions.  Why a light scalar? Does it couple just to neutrinos
(or just to $\nu_{e}$)?
Does such a light scalar fit into any reasonable unification scheme?
\item Can the postulated light scalar be emitted during ordinary weak processes? In particular,
does the emission of a light scalar lead to a measurable distortion of the spectra, apparent
lack of energy-momentum conservation {\em etc\/.} Our guess is that
the lack of distortion of $\beta$ spectra and violation of energy--momentum conservation
can pose a lower limit on the mass of the scalar and/or an upper limit on the strength
of its coupling.
\end{itemize}
It is rather obvious at this point that further research is needed in order to determine
the viability of the neutrino cloud model.
\section{Topological Defects, X-particles, {\em etc\/.}: the Saga of Top-Down Models.}
``Classic'' (\ie vintage $\simeq 1980$'s) grand unified theories (GUTs) contain quite a few
candidates of particles of masses around the GUT scale, ranging between
(approximately) $10^{11}$GeV to $10^{16}$GeV.
Those are supposed to have been produced -- in the form of
 leptoquarks,
{\em etc\/.} -- soon after the Big Bang, typically right after inflation. If sufficiently
stable, they may be around us and decay, in part, into UHE protons, thus explaining
the trans-GZK cosmic ray flux.  This mechanism was originally conjectured by the Chicago
group of astroparticle physicists (then including the late David Schramm) and the developments
are summarized in the comprehensive review paper of Bhattacharjee and Sigl,~\cite{bhatta2000}.
For this reason, we can restrict ourselves to a very brief summary of the situation.

In essence, {\em if} --- and that is an important {\em if} --- X-particles of any kind 
can be made
sufficiently stable to survive until the present epoch, their decay products can supply
the highest energy protons responsible for the trans-GZK cosmic rays.\footnote{A model
for (meta-)stabilizing superheavy particles has been proposed by Ellis \etal \cite{ellis}. In
that model  metastability is accomplished because the superheavy particles decay {\em via}
higher dimensional operators.}
 An interesting 
variation on this theme includes a recent article by Blasi~\etal~\cite{blasi} , in which, 
besides decay,
annihilation of superheavy particles is also considered. In essence, those authors conclude
that an annihilation cross section of the order of the usual strong interaction cross sections
(10 to a 100 mb) is needed in the annihilation scheme.
 Alternatively, a lifetime of $\simeq 10^{20}$yr is needed if
the UHECR are  to be produced through the decay of some X-particle.
There is no known (weakly broken) symmetry to protect the X-particle from
decaying rapidly.  If this mechanism is to be a viable one, it is
likely that the long decay lifetime is due to  a higher dimensional operator
as proposed in ref.~\cite{ellis}.
 A recent detailed account of this approach 
can be found in ref.~\cite{sarkartoldra} with appropriate references to earlier work.

The papers by Fodor and Katz~\cite{fodorkatz} put an interesting twist on the X--particle
saga. Instead of guessing on the basis of various theoretical considerations what the mass
of the X-particle might be, they attempt to fit the available data by letting the mass of the
X-particle float and be determined by the fitting procedure. In this way, they end up
with a mass of  $M_{X}\approx 10^{14.6}$GeV iun contrast to
the value, $M_{X}\approx 10^{12}$GeV used in ref.~\cite{ellis}. Other hypotheses, 
for instance a heavy leptoquark decaying into a quark and lepton would further increase 
the fitted $M_{X}$.  Fodor and Katz point out that
their fitted mass value leads to no contradiction with the observed X-ray and soft gamma ray
background. However, should they consider other X-particles (like the leptoquark just
mentioned) they {\em may} run into trouble with the EGRET results. A careful discussion of this
question is given  in the review of Bhattarcharjee and Sigl already cited. 
Fodor and Katz 
also give $\chi^{2}$ values for a variety of models including the standard model (SM)  or minimal
SUSY standard model (MSSM) by placing the X-particles into our own galaxy or making them 
extragalactic. Their best fit is either an extragalactic SM or MSSM X-particle. However,
 it may be prudent to await the arrival of some more data before
such details can be extracted.
 
Another worry is based on the result of ref.~\cite{stanevetal}. As mentioned before, 
those authors find a geodesic GZK distance about a factor of 4 or so than previously
believed. Hence, the density of X-particles has to be increased by a factor of about 45.
{\em Does this not overclose the Universe?} However, Fodor and Katz assume a
uniform distribution of X-particles. In reality, however,
X-particles cluster around galaxies, due to the gravitational attraction of
the latter.  Hence, the danger of overclosing the
Universe is, perhaps,  avoided.

 Fodor and Katz obtain
an X particle mass which is, roughly, two orders of magnitude smaller than the ``canonical''
 SUSY GUT mass, but it is about two orders of magnitude larger than a possible intermediate
mass scale of the order of $10^{12}$GeV, occurring in some SUGRA models.
Normally, it is hard to maintain mass differences of this magnitude in a
theory unless the small mass is protected by some symmetry. No such symmetry is known.
It is unclear how the hierarchy problem raised by the result of Fodor and Katz can
be avoided.

\section{A String Inspired Model.}
It has been realized a few years ago that, internal consistency of  string theories 
requires that strings live in a multidimensional space (typically, d=10 for superstrings),
the connection between a string scale and the Planck scale is less rigid than hitherto
believed~\cite{lykken, antoniadis, dienes}. Although the original conjecture of TeV scale
quantum gravity and millimeter  size compactified extra dimensions is all but excluded now,
it is useful to keep in mind such models as a paradigm from which useful features can be
abstracted.  At present, {\em there is no internally consistent, 
phenomenologically viable string model known in which even the basic features of the
dynamics -- including a mechanism of compactification -- would be satisfactorily understood.}

Nevertheless, various string models have so many attractive properties  that one is tempted
to abstract their robust features and see whether some reasonable conjectures can be made
once CMS energies of colliding particles reach the string scale.
 For the sake of argument,
let us have a string scale of the order of a 100TeV in mind. This can be reached in UHE
cosmic ray interactions: for instance, the ``gold plated'' Flye's Eye event has about
600TeV in the CMS.

Basically, two new phenomena are beginning to be observed around the string scale.
\begin{itemize}
\item A large number of excitations begins to show up with, presumably, increasing widths.
As a consequence, at least some cross sections exhibit a rapid rise towards a value which
saturates unitarity. The excitations are, in essence, of two types.
\begin{enumerate}
\item {\em Kaluza-Klein (K-K) type excitations} if the extra dimensions are compactified. Similar
excitations may take place if we live on a brane and the extra dimensions are not 
compactified, see~\cite{randallsundrum}: there are brane fluctuations.
The common feature of these excitations is that their level density grows like a power
of the CMS energy; hence, at best, a cross section (for instance the $\nu N$ cross section)
can grow only as a power of the energy. This is inadequate for explaining the 
trans-GZK cosmic rays, see Kachelriess and Pl\"{u}macher\cite{kachel}.
\item {\em The string excitations.} It is widely known that the density of states in the
excitation spectrum of strings grows asymptotically as $\exp(a \sqrt{s})$ for
$s\gg M_{s}^2$,  see, for instance, Polchinski's~\cite{polchin} book on string theory.
It is less widely known that the low lying excitations exhibit a more rapid rise of the density
of states, typically, $\propto \exp( b s)$, where $b$ is some model dependent constant. 
(Unlike the constant $a$,  related to the central charges of the underlying conformal
field theory, at present we do not know how to interpret the constant $b$ in terms
of the specific conformal field theory  considered.)
\end{enumerate}
\item There appears to be a unification of interactions. In fact, as far as one can tell,
once the energy gets into the string regime (after the first few excitations), all couplings 
are the same. There remain questions, of course, for instance, ``{\em is the unification 
mass equal to
the characteristic string mass, $M_{s}$?} Perhaps there are factors of order unity between the
two scales.  We assume that they are the same, for lack of a really
good model.
\end{itemize}

From the point of view of guessing a useful phenomenology, one conjectures that, in essence,
there are three regimes of the future theory.
\begin{enumerate}
\item {\em The low energy regime} characterized by the fact that coefficients of non 
renormalizable operators in an effective field theory, proportional to some
positive power of $s/M_{s}^{2}$ are small.
\item {\em The transition regime}, in which K-K excitations drive the coupling constants towards
a common value, as discussed, {\em e.g.} by Dienes,~ref.~\cite{goteborg}
\item {\em The string regime proper}, with all interactions unified and cross sections, in
essence governed by their unitary limits.
\end{enumerate}

It was pointed out some time ago~\cite{domonuss}  that neutrinos could be ideal primaries of
the trans-GZK cosmic rays since they have an essentially infinite mean free path in the CMBR.
If one can arrange for a stronger than SM interaction with air nuclei, one could perhaps solve
the puzzle of the trans-GZK events. The scenario sketched above provides the 
appropriate mechanism. In collisions with a CMBR photon, a neutrino of, say $E=10^{21}$eV,
has typically $\sqrt{s}\approx 200 $MeV or so: this is deep in the low energy regime even
of the SM. By contrast, in interactions with a nucleon in an air nucleus, one has
roughly $\sqrt{s}\approx 10^{3}$TeV. Thus, we might  be in the regime of unified interactions 
and string excitations~\cite{prl1}. 

Unlike some other scenarios, this one is predictive. One can easily understand in
qualitative terms one of the most robust predictions~\cite{jhep1}. It is known that
in typical GUT schemes (the GUT group always containing 
$ SU(3)_{{\rm c}}\otimes SU(2)_{{\rm L}}\otimes U(1)_{Y}$), the dominant decay mode of the
leptoquark excited in a $\nu$ nucleon interaction is a lepton and a quark. Therefore,
the shower starts as if one had a lepton induced and a hadron induced shower running
parallel to each other. In the unified regime the shower tends to be enriched in quarks,
due to the fact that quarks carry 3 times as many degrees of freedom as leptons do.
However, the energy gets spread over many particles early in the development of the
shower and, consequently, most of the evolution is governed by SM physics. Due to the
fact that the leptonic component develops by means of low multiplicity 
interactions\footnote{Counting electrons and photons on the same footing, the
average multiplicity in a lepton - nucleus interaction is very close to 2.}, the
fluctuations are larger in a neutrino induced ``anomalous'' shower than in
a proton (or nucleus) induced one.

\begin{figure}[tb]
\centering
\epsfig{figure=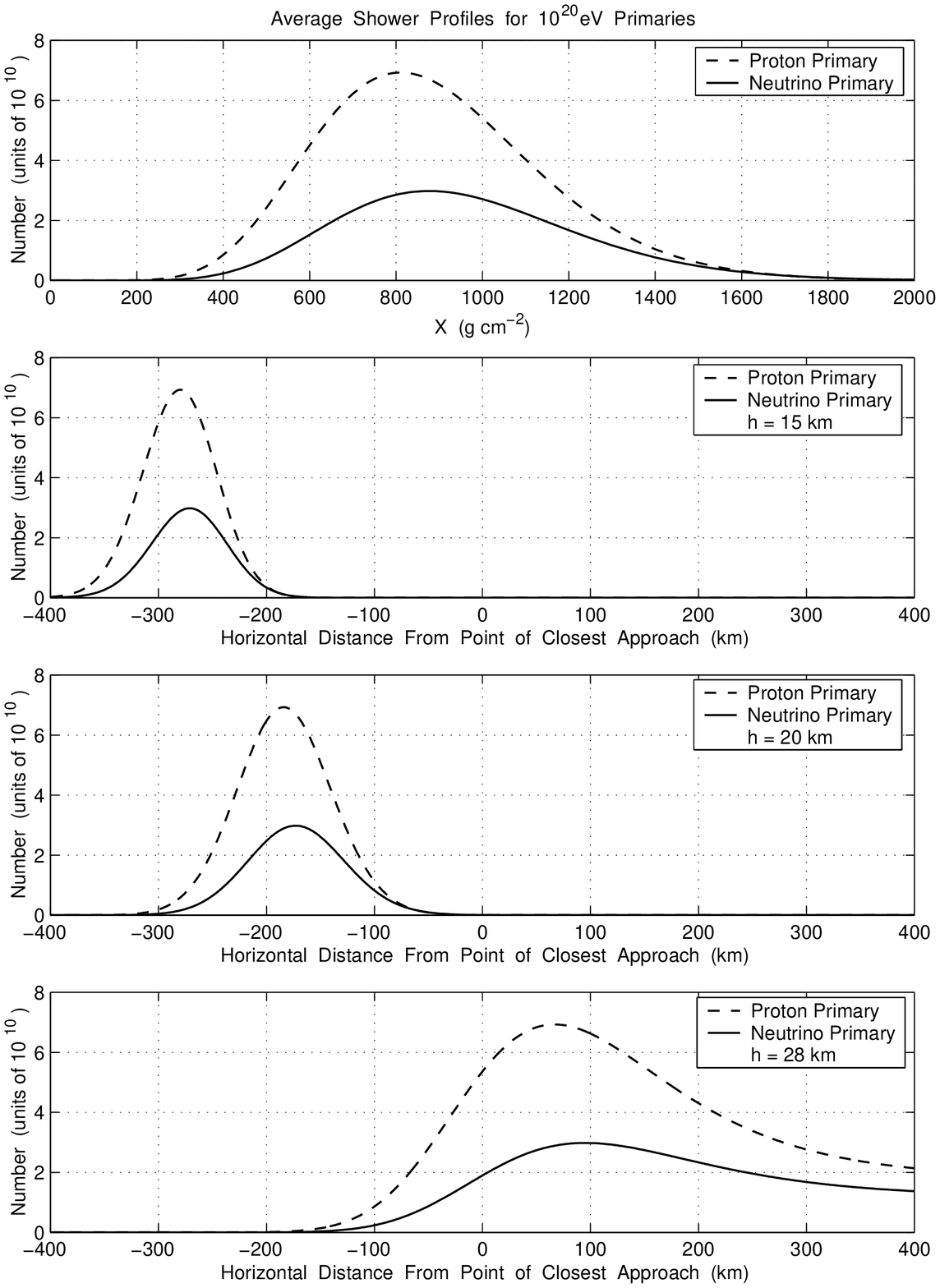, width=0.9\textwidth}
\caption{Comparison of average shower profiles for proton and neutrino 
induced showers as a function of column density and as a function of height
above the Earth for a neutrino interaction cross section equal to half
the proton value.}
\label{avgprof}
\end{figure}

In order to obtain a quantitative handle on this, the 
ALPS\footnote{Adaptive Longitudinal Profile Simulation} Monte Carlo 
 algorithm~\cite{pault} was run both for proton induced showers and for ``anomalous'' ones.
The anomalous showers were modelled along the lines just sketched; more details
can be found in ref.~\cite{jhep1}. Figure~\ref{avgprof} displays the average profiles
of a few showers with different impact parameters as it is suitable for orbiting
detectors. (Recall that the impact parameter is given by $b = R_{\oplus} +h$, where $h$
stands for the height above the surface of the Earth.) In the example shown, $M_{s}=30$TeV
was chosen. Conservatively, we assumed a $\nu$-nucleon cross section 1/2 of a hadronic one.

One could ``tweak'' the various parameters in order  to bring the $p$-induced and ``anomalous''
profiles; here we chose not to do so. However,  the qualitative properties are clear.

Most importantly, the shapes of the shower profiles are quite similar to each other. 
(This is good,
otherwise we would already know that our scheme cannot explain the trans-GZK events.)
The number of electrons at shower maximum appears to be smaller than in a proton induced
shower. By adjusting some parameters, that difference can be made smaller. In any case,
due to fluctuations, it is hard to tell on an event-by event basis what the number of
electrons at $X_{max}$ should be.

\begin{figure}[tb]
\centering
\epsfig{figure=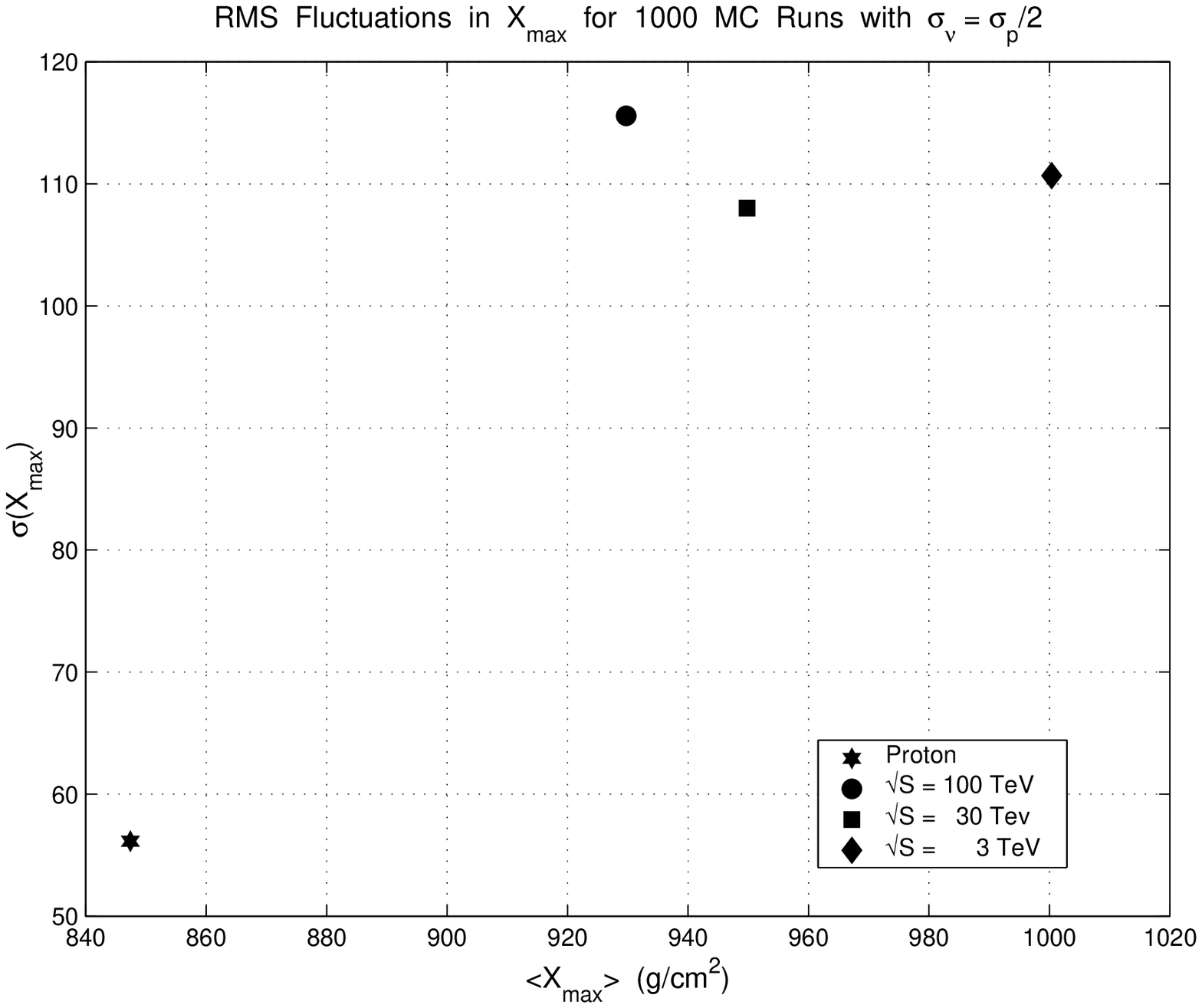,width=0.9\textwidth}
\caption{Standard deviations of the $X_{max}$ distributions shown in 
Figure\ref{standarddeviation}
for proton and neutrino induced 
showers at various values of $\sqrt{S}$ and for a primary energy $E_{0} = 10^{20}$ eV.}
\label{standarddeviation}
\end{figure}

In Figure~\ref{standarddeviation} we display the rms deviation of $X_{max}$ for both proton
induced and ``anomalous'' showers.
The result of the simulation clearly displays the feature we just discussed in qualitative terms:
``anomalous'' showers  have larger fluctuations in the electron number, hence, given sufficient
statistics, a distinction can be made between proton induced showers (presumably, 
coming from nearby) and neutrino induced ones.
\section{Discussion.}
At present, the situation is delightfully confusing, to some measure due to the unhealthy
ratio of theoretical papers to trans-GZK events observed. The situation will change
drastically in the next few years. HiRes and AGASA will continue observing.  The Pierre 
Auger observatory and the planned 
orbiting detectors will be functioning.
Hence,  the event rate of trans-GZK showers is expected to be
in the thousands per year. From the theoretical point of view, 
 all scenarios  submitted to this meeting
(and others not submitted) have some advantages and disadvantages and various
degrees of falsifiability (in Popper's sense). 
Perhaps the scheme
discussed last represents the most radical departure from established Standard Model
physics. At the same time, it has the highest degree of falsifiability. The obvious
disadvantage of that scheme is that, at present, calculations are very difficult:
as shown in an elegantly simple paper by Cornet~{\em et al}~\cite{cornet}, weakly coupled string
theories in the tree  approximation are unlikely to explain trans-GZK phenomena.
Most probably, we'll  have to learn how to handle strongly coupled string theories --- or
whatever will supercede them.

Let us end with an optimistic conclusion. We may have seen hints at physics beyond the
Standard Model or at least, some interesting new astrophysics. The jury is still out
on what the correct explanation is. In the meantime,  we have a lot of work to do.

We thank D.~Fargion, B.~McKellar, S.~Sarkar and T.~Weiler for useful comments and discussions.
 
\end{document}